\documentclass[aps,floatfix,prd]{revtex4}
\usepackage{mathtools}
\usepackage{amsmath,amssymb,epsfig,bm}
\usepackage[english] {babel}
 \addto{\captionsenglish}{}
\usepackage{amsfonts}
\usepackage{xcolor}
\usepackage{caption}
\usepackage{titlesec}
\usepackage{listings}
\usepackage{placeins}
\usepackage{graphics}
\usepackage{graphicx}
\usepackage{wrapfig}
\usepackage{etoolbox}
\usepackage{xcolor}

\usepackage{placeins}
\usepackage{mathrsfs}
\usepackage{setspace}
\usepackage[utf8]{inputenc}
\usepackage[titletoc,toc,title]{appendix}
\usepackage{wrapfig}
\usepackage{ dsfont }
\usepackage{subfig}
\usepackage{graphicx}
\usepackage{booktabs}
\numberwithin{equation}{section}
\titleformat{\paragraph}
{\normalfont\normalsize\bfseries}{\theparagraph}{1em}{}
\titlespacing*{\paragraph}
{0pt}{3.25ex plus 1ex minus .2ex}{1.5ex plus .2ex}

\newcommand{\Par}[2]{\frac{\partial {#1}  }{ \partial {#2} } }
\newcommand{\bg}{ {\text{\tiny{0}}} }

\begin{document}

\title{The Rayleigh shearing instability limit of the magnetorotational instability}
\author{Konstantinos Palapanidis$^{1}$ and Despoina Pazouli$^{1}$}
\affiliation{
$^{1}$Department of Physics, International Hellenic University, Kavala 65404, Greece}
\date{\today}

\begin{abstract}
We use the geometric optics approximation to derive the stability criteria for the Rayleigh shearing instability and the magnetorotational instability. We examine the cases where each criterion is relevant by looking into the magnitude of the magnetic field using a small dimensionless parameter. Examining all the orders of this parameter in the characteristic equation we show that configurations with sufficiently small magnetic field are characterised by the Rayleigh shearing instability criterion rather than that of the magnetorotational instability.
\end{abstract}

\maketitle

\pagenumbering{arabic}
\section{Introduction}

The dynamics of systems featuring differential rotation, like accretion discs, is a fundamental challenge in astrophysics. Its significance extends to various phenomena, including the formation of celestial bodies such as planets and stars. The magnetorotational instability (MRI) is widely accepted as a vital mechanism for elucidating the dynamics of these discs. On the other hand, purely hydrodynamic instabilities, such as the Rayleigh shearing instability, have been explored as alternative explanations, although they are not as effective as the magnetorotational instability in capturing the underlying dynamics \citep{Pringle}. 

The Rayleigh shearing instability arises from the shear in the rotation of a fluid, and has been studied extensively. It was first introduced by Lord Rayleigh in 1880, and has since been shown to play an important role in the dynamics of a wide range of fluid systems \citep{Pringle1981}. The simplest manifestation of this instability arises in an axisymmetric configuration with circular fluid motion around the axis. In this case the shear of the fluid simplifies to the radial rate of change of the angular frequency. The instability is characterised by the Rayleigh criterion. More specifically the presence of shear is a necessary but not sufficient condition for this instability to arise. 

The magnetorotational instability was first probed by Chandrasekhar  \citep{Chandra1960} and later discovered and described in its present form by Balbus \citep{Balbus1991, Balbusbook}. The MRI implies that a differentially rotating fluid, for example accretion discs around neutron stars and protoplanetary discs around young stars, is stable only if the angular velocity profile of the fluid is radially increasing, even in the case where the magnetic field is almost zero. Realistic shearing flows of astrophysical relevance have in general radially decreasing angular velocity profiles. Since most of them possess at least some very small magnetic field, they should be therefore unstable. There is a peculiarity in this result since the stability of a purely hydrodynamical system i.e. without a magnetic field, is characterised by the Rayleigh shearing instability criterion, which implies that the above mentioned velocity profiles should be stable \citep{Pringle1981}. In particular, although for most of angular velocity profiles the MRI and the Rayleigh criteria agree on the characterisation of stability, there is a set of angular velocity profiles that are characterised stable with respect to the Rayleigh criterion, but unstable with respect to the MRI criterion. These correspond to the cases where an arbitrarily small magnetic field is present in the system. 

From a physical point of view there should be no difference in the results of the two different descriptions of the same physical configuration. Rather, one would anticipate that the vanishing magnetic field limit of the MRI would provide the same results as the purely hydrodynamical Rayleigh shearing consideration. There is much discussion on this physical paradox, including a mechanical analog discussed in \cite{Balbusbook} and an allegorical analogy, written as a side note in \cite[p. 171]{Pringle} and \cite{Berry}. This analogy states that if, after taking a bite of a maggot-infested apple, you find part of the maggot, then the more maggot you find in the piece the better it is (since you ate less of the maggot). Eventually the worst case scenario is to find an infinitesimal part of the maggot (which should correspond to the case of a no-maggot-infested apple). Intuitively, this is the opposite of what one would expect, i.e. that the best case scenario is to not find any maggot in the apple at all. In analogy, strong magnetic fields provide more stability than weaker magnetic fields, which destabilise the system the weaker they are up to the limit that there is no magnetic field at all.

In the present work, we aim to discuss this paradox that appears to exist between the stability results obtained in the low magnetic field limit of magnetorotational systems and purely hydrodynamical systems. To achieve this, we will derive the Rayleigh and MRI stability criteria using the geometric optics approach. Using this approach, we will obtain the linearly perturbed system of equations describing both the purely hydrodynamic system and the magnetohydrodynamic system. In particular, we apply the geometric optics method by determining the rate of change of the background quantities with respect to the coordinates and time, and we reach the same characteristic equation as in \cite{Balbusbook}. Contrary to the original paper we do not consider the Boussinesq approximation for the continuity equation but we rather use the full form of it.

In section \ref{perturbations} we describe the system of non-linear and  linearised equations using the geometric optics approximation assuming plane wave perturbations. In the next section we describe axisymmetric configurations of purely hydrodynamic and magnetohydrodynamic systems and derive the Rayleigh shearing instability \citep{ArmitageBook} and the magnetorotational instability in agreement with the literature.
In section \ref{Sec_rs_limit_MRI} we examine the case where the magnetic field obtains very small, close to zero, values and discuss the applicability of the Rayleigh and MRI stability criteria.

\section{Linear perturbation of the system}\label{perturbations}

In this section we apply the geometric optics approximation to the purely hydrodynamical and to the ideal magnetohydrodynamical (MHD) systems of equations.\footnote{The geometric optics approach, despite its name, is not exclusively employed for electromagnetic wave problems. Instead, it can be utilized to encompass a wider array of physical phenomena \citep{AnileBook}.}Specifically in the present work we use the two-timing method.It is called as such because we use two different parameters that control the magnitude of the quantities involved \citep{Whitham1970}. We present both of the aforementioned systems of equations and we introduce the ansatz to linearise them. Finally, by keeping only the background and the first order terms, we provide the perturbed equations.
\subsection{The system of equations} \label{Sec_syst_eq}

In this section, we describe the system of equations that we use to derive the MRI and the Rayleigh shearing instability.  
The results we derive are either in the context of hydrodynamics or ideal magnetohydrodynamics. The description of a single fluid in the Newtonian framework employs the continuity equation given by 
  \begin{equation}
  \begin{aligned}
  \Par{\rho}{t} + \rho \bm{\nabla} \cdot \bm{v} + (\bm{v} \cdot \bm{\nabla}) \rho =0,
\label{cont_newt}
\end{aligned}
\end{equation}
where $\bm{v}$ is the fluid velocity and $\rho$ is the density. Please note that contrary to the original paper \citep{Balbus1991} where the Boussinesq approximation \citep{TrittonBook}, i.e. $\bm{\nabla} \cdot \bm{v}=0$,  was considered we use the full form of the continuity equation. Consequently, we do not implicitly impose additional conditions on the background and perturbed density.  We also have the Euler (momentum conservation) equation
\begin{equation}
\begin{aligned}
 \Par{\bm{v}}{t} + (\bm{v} \cdot \bm{\nabla})  \bm{v} + \frac{1}{\rho} \bm{\nabla} P + \bm{\nabla} \Phi  \underbrace{ +\frac{1}{4 \pi \rho}\bm{B}  \times \left( \bm{\nabla} \times \bm{B} \right)}_{\text{ideal MHD Lorentz force}} =0,
\label{eul_lan_newt}
\end{aligned}
\end{equation}
where $\Phi$ is the gravitational potential, $\bm{B}$ is the magnetic field and $P$ is the pressure of the fluid. For a purely hydrodynamical system the ideal MHD Lorentz force (i.e. the under-brace term) vanishes. To describe ideal MHD systems we need to include the magnetic field induction equation,
 \begin{equation}
 \begin{aligned}
    \Par{\bm{B}}{t} - \bm{\nabla} \times \left( \bm{v} \times \bm{B} \right) =0,
\label{induction_newt_B}
\end{aligned}
\end{equation}
in our system of equations as well. As discussed in the literature, \citep{BellanBook, GoossensBook} this equation is obtained by using the Maxwell equations, assuming that the fluid is perfectly conducting. 
Please note that for pure hydrodynamic systems this equation is not required, since the magnetic field is zero \citep{Pringle}. Finally, the adiabatic condition,
  \begin{equation}
  \begin{aligned}
  \Par{\Sigma}{t}  + (\bm{v} \cdot \bm{\nabla})\Sigma =0,
\label{entropy_cons_newt}
\end{aligned}
\end{equation}
is required in both the hydrodynamics and the ideal MHD cases. In the above equation,  $\Sigma$ is the specific entropy of the fluid. We assume an adiabatic flow, which means that the entropy is conserved along the flow lines \citep{RezzollaBook}. The entropy is considered to be a function of the pressure and the density, $\Sigma= \Sigma \left( P, \rho \right)$, and serves as an equation of state for the system. Please note that under this consideration the pressure and the density are independent quantities. The speed of sound is defined through
 \begin{equation}
  \begin{aligned}
 c_{\rm s}^2 = \left. \Par{P}{\rho}\right|_{\Sigma},
\label{sw_der_newt_def}
\end{aligned}
\end{equation}
and describes the speed of propagation for acoustic perturbations \citep{RezzollaBook}. Using this definition for the speed of sound, equation (\ref{entropy_cons_newt}) in terms of $P$ and $\rho$ becomes
 \begin{equation}
  \begin{aligned}
  \Par{P}{t}  + (\bm{v} \cdot \bm{\nabla})P - c_{\rm s}^2 \left[ \Par{\rho}{t}  + (\bm{v} \cdot \bm{\nabla})\rho \right]  =0.
\label{entropy_cons_newt2}
\end{aligned}
\end{equation}

\subsection{Linear perturbations} \label{Sec_pert_newt}

In this section we calculate the linear perturbations of the system of equations of section (\ref{Sec_syst_eq}) using the geometric optics approximation. We substitute all quantities of the system using the ansatz
\begin{equation}
\begin{aligned}
  \rho = \rho_\bg + \delta \rho,
 \label{linearisation}
 \end{aligned}
\end{equation} 
where $\rho_\bg$ is a background quantity and
\begin{equation}
\begin{aligned}
 \delta \rho= \bar{\delta}\: {\rm e}^{{\rm i} \frac{S}{\bar{\varepsilon}}}\: \bar{\rho}\left(\bar{\varepsilon} t, \bar{\varepsilon} \bm{r} \right) ,
 \label{ansatz}
 \end{aligned}
\end{equation} 
is the linear perturbation of $\rho$, which describes a locally plane wave with amplitude $\bar{\rho}$ and phase $S$ by definition \citep{AnileBook, PerlickBook, WhithamBook}. The quantities $\bar{\delta}$ and $\bar{\varepsilon}$ are small ($0<\bar{\varepsilon}<\bar{\delta}\ll 1$) dimensionless book-keeping parameters used to keep track of the ordering of the linearised terms. In particular we keep only terms of the order $\bar{\delta}^0 \bar{\varepsilon}^0$ and $\bar{\delta}^1 \bar{\varepsilon}^0$ which are the background and the linearised terms respectively. Higher order terms in $\bar{\delta}$ are disregarded since they are higher order perturbation terms. Similarly, higher than zeroth order in $\bar{\varepsilon}$ terms are also not considered since they correspond to post-geometric optics approximations \citep{AnileBook}. The quantity $\bar{\rho}$, which is the amplitude of the perturbation, is assumed to be of the order of unity, while $S$, which is the phase of the plane wave \citep{BornOptics}, is given by the equation
\begin{equation}
\begin{aligned}
 S=\bar{\varepsilon} ( \bm{k}\cdot \bm{r} - \omega t),
 \label{phase_newt}
 \end{aligned}
\end{equation} 
where $\bm{k}$ is the wavevector, $\bm{r}$ is the position vector, and $\omega$ is the frequency. Please note that we use the ansatz presented in equations \ref{linearisation}-\ref{ansatz} and the expressions developed above for the linearisation of all the quantities involved in the equations except for the gravitational potential, where we have assumed that $\delta \Phi =0$, which is assumed to be a background quantity only.  

The background terms (i.e. those of order $\bar{\delta}^0\bar{\varepsilon}^0$) satisfy the system of equations (\ref{cont_newt})-(\ref{induction_newt_B}), (\ref{entropy_cons_newt2}) and therefore vanish identically, and the first-order terms are the only to appear in the linearised equations. We find that the continuity equation (\ref{cont_newt}) in its linearised form (i.e. containing terms of the order $\bar{\delta}^1 \bar{\varepsilon}^0$) is given by 
 \begin{equation} 
 \begin{aligned}
&- {\rm i\,} \omega \bar{\rho} +\bar{\rho} \left( \bm{\nabla} \cdot \bm{v}_\bg \right) + {\rm i\,} \rho_\bg \left( \bm{k} \cdot \bar{\bm{v}} \right) + \rho_\bg \left( \bm{\nabla} \cdot  \bar{\bm{v}}  \right)\\
& + {\rm i} \left( \bm{v}_\bg \cdot \bm{k} \right) \bar{\rho} + \bar{\bm{v}} \cdot \bm{\nabla} \rho_\bg =0.
\label{cont_pert_newt}
\end{aligned}
\end{equation}
Similarly, the Euler equation (\ref{eul_lan_newt}) obtains the following form
\begin{equation}
 \begin{aligned}
 & - {\rm i\,} \omega \bar{\bm{v}} + \left( \bar{\bm{v}} \cdot \bm{\nabla} \right) \bm{v}_\bg +{\rm i\,} \left( \bm{v}_\bg \cdot \bm{k} \right) \bar{\bm{v}}   + \left( \bm{v}_\bg \cdot \bm{\nabla} \right) \bar{\bm{v}}\\
 & -\frac{\bar{\rho}}{\rho_\bg^2} \bm{\nabla} P_\bg + {\rm i\,} \frac{\bm{k}}{\rho_\bg} \bar{P}  -\frac{\bar{\rho}}{4 \pi \rho_\bg^2} \bm{B}_\bg \times \left( \bm{\nabla} \times \bm{B}_\bg \right) 
\\
& +\frac{1}{4 \pi \rho_\bg}\left[ \bar{\bm{B}} \times \left( \bm{\nabla} \times \bm{B}_\bg \right) +  \bm{B}_\bg \times \left( \bm{\nabla} \times \bar{\bm{B}} \right)\right]  =0,
 \end{aligned}
\label{eul_lan_newt_pert_B}
\end{equation}
where the terms containing the magnetic field is the linearised ideal MHD Lorentz force.
The induction equation (\ref{induction_newt_B}) becomes
 \begin{equation}
  \begin{aligned}
  &  -{\rm i\,} \omega \bar{\bm{B}} + \bar{\bm{B}} \left( \bm{\nabla} \cdot \bm{v}_\bg \right) + {\rm i\,} \bm{B}_\bg \left( \bm{k} \cdot \bar{\bm{v}} \right)\\
    &     - \left( \bar{\bm{B}} \cdot \bm{\nabla} \right) \bm{v}_\bg - \left( \bm{B}_\bg \cdot \bm{\nabla} \right) \bar{\bm{v}} + \left( \bm{v}_\bg \cdot \bm{\nabla} \right) \bar{\bm{B}} =0.
\label{induction_newt_pert_B}
\end{aligned}
 \end{equation}

Finally, the adiabatic condition (\ref{entropy_cons_newt}) yields
 \begin{equation} 
    \begin{aligned}
      &  {\rm i} \left( \bm{v}_\bg \cdot \bm{k} - \omega \right) \left(  \bar{P} - c_{\rm s}^2 \bar{\rho}  \right) + \bar{\bm{v}} \cdot \bm{\nabla} P_\bg -c_{\rm s}^2 \,   \bar{\bm{v}} \cdot \bm{\nabla} \rho_\bg \\
    & - \bar{\rho} \left(  \left.\Par{c_{\rm s}^2}{P}\right|_{\rho_\bg} \bm{v}_\bg \cdot \bm{\nabla} P_\bg  + \left.\Par{c_{\rm s}^2}{\rho}\right|_{P_\bg} \bm{v}_\bg \cdot \bm{\nabla} \rho_\bg    \right)\\
    &+  \frac{ \bar{P} - c_{\rm s}^2 \bar{\rho} }{ \left. \partial \Sigma_\bg / \partial P \right|_{\rho_\bg}}\left( \left.\frac{\partial^2 \Sigma_\bg}{\partial P^2}\right|_{\rho_\bg} \right) \bm{v}_\bg \cdot \bm{\nabla} P_\bg =0,
\label{entr_cons_pert_newt}
\end{aligned}
\end{equation}
where we used the assumption that the specific entropy is a function of pressure and density along with the definition (\ref{sw_der_newt_def}) of the speed of sound. For a detailed derivation of the above equation see \cite{KonPhD}.

\section{Axisymmetric configurations}

In this section we discuss two axisymmetric configurations, the Rayleigh shearing instability and the magnetorotational instability. The Rayleigh shearing instability \citep{Pringle1981} characterizes the stability of a purely hydrodynamic system, i.e without a magnetic field. It is a well known result that the stability criterion depends on the angular velocity profile. In contrast, the MRI \citep{Balbus1991, Balbusbook, Balbus2016} is characterized through the angular velocity profile of systems in the context of ideal MHD.

Since the physical systems of interest mainly are disc-shaped configurations around astrophysical objects, such as accretion and protoplanetary discs around stars, we will carry out our calculations in cylindrical polar coordinates $\left( R,z,\phi \right)$ and the respective orthonormal frame $\hat{\bm{R}},\hat{\bm{z}},\hat{\bm{\phi}}$. In all cases we assume that all quantities are axisymmetric, i.e. they do not depend on the $\phi$ coordinate (though, we may still have vector components along $\hat{\bm{\phi}}$). Additionally, since we wish to demonstrate that the vanishing magnetic field limit of the MRI is the Rayleigh shearing instability, we make the same assumptions for the quantities involved in the derivation of both stability criteria.

\subsection{The Rayleigh shearing instability}
\label{Rayleigh_section}

In order to introduce the Rayleigh shearing instability, we will first describe the equilibrium of the system. 
 We assume that the system consists of a fluid that is differentially rotating around the $z$-axis, having a velocity of the form $\bm{v}_\bg :=\Omega(R) R \hat{\bm{\phi}}$, where $\Omega$ is the angular velocity of the fluid. The density, pressure and gravitational potential have the functional forms, $\rho:=\rho_\bg\left(\bar{\varepsilon} t, \bar{\varepsilon} R, \bar{\varepsilon} z \right)$, $P_\bg:=P_\bg\left(\bar{\varepsilon} t, R,z\right)$, and $\Phi_\bg:=\Phi\left( \bar{\varepsilon} t, R,z \right)$ respectively. Our assumptions describe a system where the background density varies slowly in any direction and in time, and the background pressure and gravitational potential varies slowly in time only, but have fast space dependence along both $(R,z)$ coordinates. Also, the angular velocity of the fluid is fast along the radial coordinate R. Please note that by considering dependence of the form $\bar{\varepsilon}z$ along the $z$ coordinate, as defined above, we are introducing a very small variation of the respective quantity along the $z$ direction. This is obvious if we consider the chain rule of differentiation, where the derivative with respect to $z$ is multiplied by $\bar{\varepsilon}$. The same holds for the rest of the coordinates and time.
 
Using these assumptions, and keeping terms of order $\bar{\delta}^0 \bar{\epsilon}^0$ only, we derive the hydrostatic equilibrium condition given by the $R$ and $z$ components of the Euler equation (\ref{eul_lan_newt})
\begin{equation}
   \Omega^2\, R = \Par{\Phi_\bg}{R} + \frac{1}{\rho_\bg} \Par{P_\bg}{R},
\label{rs_cyl_eul_R_bg}
\end{equation}
and
 \begin{equation}
   \Par{\Phi_\bg}{z} + \frac{1}{\rho_\bg} \Par{P_\bg}{z}=0,
\label{rs_cyl_eul_z_bg}
\end{equation}
respectively.
The equations above describe a fluid in stationary equilibrium where the gravitational force along the radial direction is balanced by the gradient of the pressure and the centripetal force, while in the $z$ direction we only have the pressure gradient and the gravitational force. The rest of background equations (\ref{cont_newt}) and (\ref{entropy_cons_newt2}) are trivially satisfied for the assumptions we have made. We will consider wavevectors of the form $\bm{k}=k_R \hat{\bm{R}} + k_z \hat{\bm{z}}$. This wavevector describes a plane wave that has no $\phi$ dependence. Our choice is justified by the fact that the system we assume is axisymmetric.
the linearised continuity equation (\ref{cont_pert_newt}) is given by 
 \begin{equation}
\rho_\bg \left( \frac{1}{R_\bg}+ {\rm i\,} k_R \right) \bar{v}_R + {\rm i\,} \rho_\bg k_z \bar{v}_z - {\rm i\,} \omega \bar{\rho} =0,
\label{sw_cyl_cont_newt}
\end{equation}
while the $R$, $z$ and $\phi$ components of the linearised Euler equation (\ref{eul_lan_newt_pert_B}) are
 \begin{equation}
-{\rm i\,} \omega \bar{v}_R - 2 \Omega \bar{v}_\phi -\frac{1}{\rho_\bg^2} \Par{P_\bg}{R} \bar{\rho} + {\rm i\,} \frac{k_R}{\rho_\bg} \bar{P} =0, 
\label{rs_cyl_eul_newt_R}
\end{equation}
 \begin{equation}
 -{\rm i\,} \omega \bar{v}_z -\frac{1}{\rho_\bg^2} \Par{P_\bg}{z}  \bar{\rho}  + {\rm i\,} \frac{k_z}{\rho_\bg} \bar{P} =0 ,
\label{rs_cyl_eul_newt_z}
\end{equation}
and
 \begin{equation}
\left(  2 \Omega + R \frac{{\rm d} \Omega}{{\rm d}R}  \right)  \bar{v}_R -{\rm i\,} \omega \bar{v}_\phi =0,
\label{rs_cyl_eul_newt_phi}
\end{equation}
respectively. As we have already mentioned previously, in this purely hydrodynamic system the magnetic field terms and the linearised induction equation (\ref{induction_newt_pert_B}) are omitted. The linearised adiabatic condition (\ref{entr_cons_pert_newt}) takes the form
 \begin{equation} 
  \begin{aligned}
  \Par{P_\bg}{R} \bar{v}_R  +  \Par{P_\bg}{z}  \bar{v}_z   - {\rm i\,} \omega \bar{P} + {\rm i\,} \omega c_{\rm s}^2 \bar{\rho} =0.
\label{rs_newt_pert_entrocons}
\end{aligned}
\end{equation}
Please note that the  directional derivatives $\bm{v}_\bg \cdot \bm{\nabla}$ of scalars are vanishing, as in equation (\ref{entr_cons_pert_newt}). This happens because the fluid velocity has a single component along $\hat{\bm{\phi}}$ and axisymmetric quantities do not have a $\phi$ dependence. Equations (\ref{sw_cyl_cont_newt}--\ref{rs_newt_pert_entrocons}) comprise a system of five equations in five variables. The variables are the perturbation amplitudes of the physical quantities, namely $\bar{\rho},\: \bar{P},\: \bar{v}_R,\: \bar{v}_z,\: \bar{v}_\phi$. This system has a non-trivial solution if and only if the determinant of the matrix of the coefficients of the variables is zero \citep{SwansonBook}. The vanishing of the determinant yields the characteristic equation (along with the $\omega=0$ solution) of the system
  \begin{equation}
\begin{aligned}
& \left( \omega^2 k^2 - k_z^2 \kappa^2 \right) c_{\rm s}^2 +  \left[ -\omega^4 + \omega^2 \kappa^2 + \frac{1}{\rho_\bg^2}   \left( \Par{P_\bg}{z} k_R - \Par{P_\bg}{R} k_z   \right)^2  \right] \\
 & + \frac{1}{R} \left[ \omega^2 \left( \frac{1}{\rho_\bg} \Par{P_\bg}{R} -{\rm i\,} c_{\rm s}^2 k_R \right) + \frac{{\rm i\,}}{\rho_\bg^2} \Par{P_\bg}{z} \left( \Par{P_\bg}{R} k_z  -\Par{P_\bg}{z} k_R \right)    \right] =0,
\label{rs_newt_full_char_eq}
\end{aligned}
\end{equation}
where $\kappa$ is the epicyclic frequency \citep{Pringle} given by
  \begin{equation}
 \kappa^2 = 4 \Omega^2 +2 R \Omega \frac{{\rm d} \Omega}{{\rm d} R}.
\label{epicycl_freq}
\end{equation}
The expression given in equation (\ref{rs_newt_full_char_eq}) is further simplified using two assumptions. Firstly, we consider the cases that are not close to the axis of symmetry. A similar treatment is used in \cite{Balbusbook} in the sense that $1/R$ is of the order of $\bar{\varepsilon}^\alpha, \:\: \alpha \ge 1$. Secondly, we also eliminate the sound waves considering the case where the speed of sound is large, that is $c_{\rm s}^2$ is of the order of $\bar{\varepsilon}^{\beta}, \:\; \beta \le -1$, and diving the equation by $c_{\rm s}^2$. Since in our analysis we only keep zero order terms in $\bar{\varepsilon}$, the surviving terms yield
  \begin{equation}
\omega^2  =  \kappa^2 \frac{k_z^2}{k^2}.
\label{rs_newt_red_char_eq}
\end{equation}
The system is stable if, for real values of $k$, $\omega$ is real. This in turn means that for stability the condition 
\begin{equation}
 \kappa^2 \ge 0
 \label{RS_criterion}
\end{equation} 
holds. If this criterion is not satisfied it gives rise to the well known Rayleigh shearing instability \citep{Pringle, Balbusbook}. Combining the definition for the epicyclic frequency (\ref{epicycl_freq}) with the above criterion we obtain the inequality
\begin{equation}
  \Omega\, \frac{{\rm d}}{{\rm d}R}\left(\Omega R^2 \right) \ge 0,
 \label{RS_criterion_easy}
\end{equation} 
which is satisfied if $\Omega>0$ and $\Omega R^2$ is an increasing function of R, or if $\Omega<0$ and $\Omega R^2$ is a decreasing function of R. Note that these two cases are equivalent since the sign of $\Omega$ is a matter of convention. Also note that in both cases the absolute value of the differentiated quantity increases with $R$ when the criterion is satisfied, as expected. This inequality means that even if $\Omega$ is a decreasing function of $R$, the system is still stable as long as $\Omega R^2$ is increasing along $R$.
 
\subsection{The magnetorotational instability }
\label{MRI}

In this section we derive the magnetorotational instability \citep{Balbus1991, Balbusbook}. This is done by introducing the magnetic field in the configuration of the previous section. The background quantities for the fluid density, the pressure, and the velocity, as well as the wavevector are the same as in the Rayleigh shearing configuration. For the magnetic field we consider components only along the $z$ and the $\phi$ direction, $\bm{B}_\bg=B_{\bg,z}\left(\bar{\varepsilon}t,\bar{\varepsilon} R, \bar{\varepsilon} z \right)\hat{\bm{z}}+B_{\bg,\phi}\left(\bar{\varepsilon}t,\bar{\varepsilon} R, \bar{\varepsilon} z \right)\hat{\bm{\phi}}$. As mentioned previously the main difference between this analysis and Balbus' original paper \citep{Balbus1991} is that we use the full continuity equation instead of the Boussinesq approximation. Also we avoid the assumption that isobaric and isochoric surfaces coincide which may be somewhat restrictive. 

The background equations are, as in the Rayleigh shearing instability case, the $R$ and $z$ components of the Euler equation  (\ref{eul_lan_newt}). The $z$ component is given by equation (\ref{rs_cyl_eul_z_bg}) while for the $R$ component we have
  \begin{equation}
   \Omega^2\, R = \Par{\Phi_\bg}{R} + \frac{1}{\rho_\bg} \Par{P_\bg}{R} + \frac{B^2_{\bg,\phi}}{R},
\label{MRI_cyl_eul_R_bg}
\end{equation}
where the extra term, compared to equation (\ref{rs_cyl_eul_R_bg}), appears due to the ideal MHD Lorentz force. The rest of the background equations vanish identically. The $R$ component of the linearised Euler equation (\ref{eul_lan_newt_pert_B}) is 
  \begin{equation}
  \begin{aligned}
&-{\rm i\,} \omega \bar{v}_R - 2 \Omega \bar{v}_\phi - \left( \frac{1}{\rho_\bg^2} \Par{P_\bg}{R}  +\frac{1}{R}\, \frac{B_{\bg,\phi}^2 k_z}{4 \pi \rho^2_\bg}
 \right) \bar{\rho} + {\rm i\,} \frac{k_R}{\rho_\bg} \bar{P} \\
&  -{\rm i\,} \frac{ B_{\bg,z} k_z}{4 \pi \rho_\bg} \bar{B}_R 
  +{\rm i\,} \frac{ B_{\bg,z} k_R}{4 \pi \rho_\bg} \bar{B}_z  + \frac{B_{\bg,\phi}}{4 \pi \rho_\bg} \left( \frac{2}{R} + {\rm i\,} k_R \right) \bar{B}_\phi   =0, 
\label{MRI_cyl_eul_newt_R}
\end{aligned}
\end{equation}
 the $z$ component is 
 \begin{equation}
 -{\rm i\,} \omega \bar{v}_z  + {\rm i\,} \frac{k_z}{\rho_\bg} \bar{P} -\frac{1}{\rho_\bg^2} \Par{P_\bg}{z} \bar{\rho} + {\rm i\,} \frac{ B_{\bg,\phi} k_z}{ 4 \pi \rho_\bg } \bar{B}_\phi =0 ,
\label{MRI_cyl_eul_newt_z}
\end{equation}
and the $\phi$ component is
 \begin{equation}
  \left(2 \Omega  + R \frac{{\rm d} \Omega}{{\rm d} R} \right) \bar{v}_R -{\rm i\,} \omega \bar{v}_\phi - \frac{1}{R} \frac{B_{\bg,\phi}}{4 \pi \rho_\bg} \bar{B}_R - {\rm i\,} \frac{B_{\bg,z} k_z}{ 4 \pi \rho_\bg} \bar{B}_\phi =0.
\label{MRI_cyl_eul_newt_phi}
\end{equation}
The components of the linearised induction equation (\ref{induction_newt_pert_B}) are
  \begin{equation}
- {\rm i\,} B_{\bg,z} k_z \bar{v}_R - {\rm i\,} \omega \bar{B}_R =0,
\label{MRI_cyl_induct_R}
\end{equation}
 \begin{equation}
  B_{\bg,z} \left( \frac{1}{R} + {\rm i\,} k_R \right) \bar{v}_R - {\rm i\,} \omega \bar{B}_z =0,
\label{MRI_cyl_induct_z}
\end{equation}
and
  \begin{equation}
   {\rm i\,} B_{\bg,\phi} k_R \bar{v}_R + {\rm i\,} B_{\bg,\phi} k_z \bar{v}_z - {\rm i\,} B_{\bg,z} k_z \bar{v}_\phi - R \frac{{\rm d} \Omega}{{\rm d} R} \bar{B}_R - {\rm i\,} \omega \bar{B}_\phi=0,
\label{MRI_cyl_induct_phi}
\end{equation}
for the $R$,$z$ and $\phi$ components respectively. Please note that the linearised continuity equation and linearised adiabatic condition are given by (\ref{sw_cyl_cont_newt}) and (\ref{rs_newt_pert_entrocons}) respectively, since these equations are independent of the magnetic field. 

This is a system with eight equations and eight unknowns, the three extra unknowns (compared to the hydrodynamic system) being the perturbations of the three components of the magnetic field. The full characteristic equation, which is a sixth degree polynomial in $\omega$, can be found in \ref{appendix}. It contains the acoustic modes and the terms related to small radial distances, which make it quite a cumbersome expression and it can not be treated analytically. Following the method used in the purely hydrodynamic case of the previous section, we eliminate the sound waves and terms of order $1/R$ or smaller. Having applied these simplifications, the characteristic equation of the system is given by
 \begin{equation}
\frac{k^2}{k_z^2} \omega^4 - \left( \kappa^2 + 2 k^2 v_{\rm Az}^2  \right)   \omega^2 +k_z^2 v_{\rm Az}^2 \left( \kappa^2 - 4 \Omega^2 + k^2 v_{\rm Az}^2 \right) =0,
\label{MRI_cyl_char_eq}
\end{equation}
where $v_{\rm Az}$ is the Alfv\'{e}n speed along the $z$ direction and $v_{\rm Az}^2=\frac{B_{\bg,z}^2}{4 \pi \rho_\bg}$. Please note that the Alfv\'{e}n speed is directly proportional to the magnetic field. In the following section we use $v_{\rm Az}^2$ to examine the low magnetic field behaviour of the system and we assume that the density is of the order of unity. This characteristic equation is identical to the one derived in \cite{Balbusbook} if we consider a wavevector with only a $z$ component, i.e. $k_R=0$.
 The left-hand-side of equation (\ref{MRI_cyl_char_eq}) is a convex quadratic polynomial in $\omega^2$, since the coefficient of $\omega^4$ is positive. The discriminant is $k_z^4 \left( \kappa^4 + 16 k^2 \Omega^2 v_{\rm Az}^2 \right)$ which is always positive and therefore the two roots of the polynomial are real. 
Additionally the two $\omega^2$ roots are positive and thus the system is stable if the coefficient of $\omega^2$ is negative  and the constant term is positive. The first condition implies that the minimum of the polynomial occurs at positive $\omega^2$ and the second condition implies that the polynomial intersects the $\omega^2=0$ axis at a positive value. These two conditions read
  \begin{equation}
  \kappa^2 + 2 k^2 v_{\rm Az}^2 \ge 0,
\label{MRI_stability_cond1}
\end{equation}
and
  \begin{equation}
 \kappa^2 - 4 \Omega^2 + k^2 v_{\rm Az}^2 \ge 0.
\label{MRI_stability_cond2}
\end{equation}
Of these two inequalities (provided that always $ k^2 v_{\rm Az}^2 >0$) we only need the second one because if it is satisfied, the first is satisfied as well.
Assuming then that $k^2 v_{\rm A z}^2$ goes to zero (since we can either have a very small magnetic field or very small wavenumbers) the stability criterion reads
 \begin{equation}
\kappa^2 \ge  4 \Omega^2.
\label{MRI_stability_cond4}
\end{equation}
We will call this the Balbus criterion. Using the definition of the epicyclic frequency from equation (\ref{epicycl_freq}) the stability condition obtains the following form
 \begin{equation}
\frac{{\rm d} \Omega^2}{{\rm d} \ln{R}}  \ge 0,
\label{MRI_stability_cond4_a}
\end{equation}
which is the one derived in \cite{Balbusbook}. A simpler form for the above is
 \begin{equation}
\Omega \frac{{\rm d} \Omega}{{\rm d}R} \ge 0,
\label{MRI_stability_cond4_b}
\end{equation}
which is satisfied if $\Omega>0$ and increasing along $R$ or, if $\Omega<0$ and decreasing along $R$. Similarly to the Rayleigh criterion (\ref{RS_criterion_easy}), in both cases the condition requires that the absolute value of $\Omega$ is increasing along $R$. In contrast to the Rayleigh configuration, a disc is stable if the magnitude of $\Omega(R)$ is radially increasing outwards. However, for most astrophysical configurations $\Omega(R)$ decreases in magnitude with respect to the radius and so the majority of realistic models should be unstable \citep{ArmitageBook, Balbusbook}. 

A peculiar and interesting aspect of this result is that the vanishing magnetic field condition (\ref{MRI_stability_cond4}) does not coincide with the Rayleigh shearing instability criterion of the previous section, as we would anticipate.
Physically this means that an arbitrarily small magnetic field would produce an instability in a configuration which would be stable if the magnetic field had not been introduced at all, i.e because we may have $\kappa^2>0$ (Rayleigh criterion) but $\kappa^2 - 4 \Omega^2<0$ (MRI criterion with vanishing magnetic field). In the following section, we will discuss the vanishing magnetic field limit of the MRI.

\section{Vanishing magnetic field limit of the MRI} \label{Sec_rs_limit_MRI}

The condition (\ref{MRI_stability_cond4}) arises by taking the limit $k^2 v_{\rm A z}^2 \rightarrow 0$. Obviously, this limit is achieved if any of $k$, $v_{\rm A z}$, or both, approach zero. In the analysis below we consider the case where the magnitude of the Alfv\'{e}n speed is controlled only by the magnitude of the magnetic field (i.e. the density is of order unity). This is reasonable since we are interested in the vanishing magnetic field limit of the MRI. We will look into condition (\ref{MRI_stability_cond2}) in more detail, by examining all the possible range of values for the quantities $k$, $v_{\rm A z}$.

 We introduce a function
 \begin{equation}
  \zeta = k^2 v_{\rm A z}^2,
\label{MRI_zeta_bookkeep}
\end{equation}
that will be used to keep track of the magnitude of this term. This quantity, being the product of $k^2$ and $v_{\rm A z}^2$, serves as the Alfv\'{e}n angular frequency squared. The characteristic equation (\ref{MRI_cyl_char_eq}) then reads
\begin{equation}
\begin{aligned}
& \cos^2 q \zeta^2 + \left( \kappa^2  \cos^2 q- 4 \Omega^2 \cos^2 q - 2  \omega^2 \right) \zeta  + \omega^2 \left( \frac{\omega^2}{\cos^2 q}  - \kappa^2 \right) =0.
\end{aligned}
\label{MRI_cyl_char_eqA1}
\end{equation}
where $\cos{q}$ is the direction cosine, defined by $k_z=k \cos{q}$. Note that we rearranged equation (\ref{MRI_cyl_char_eq}) in powers of $\zeta$. Also, note that the direction cosine does not vanish since it would imply $k_z=0$ and subsequently only $\omega=0$ solutions. There are three cases to be considered with respect to the value of $\zeta$ and the stability condition derived from equation (\ref{MRI_cyl_char_eqA1}). 

The first case corresponds to $\zeta$ large enough compared to the  $\kappa^2 - 4 \Omega^2$ term so that no term of equation (\ref{MRI_cyl_char_eqA1}) can be neglected and therefore the stability condition is given by inequality (\ref{MRI_stability_cond2}). The $k^2 v_{\rm A z}^2$ is included in the final criterion because it is of the same magnitude as the other term, as mentioned in the previous section as well. 

The second case occurs when the product of the Alfv\'{e}n speed and the wavenumber is such that the $\zeta^2$ is sufficiently small compared to the other background terms of the characteristic equation to be omitted, but $\zeta$ is not. In this case equation (\ref{MRI_cyl_char_eqA1}) reduces to
 \begin{equation}
\frac{1}{\cos^2 q } \omega^4 - \left( \kappa^2 + 2 k^2 v_{\rm Az}^2  \right)   \omega^2 + (\cos^2 q) k^2 v_{\rm Az}^2 \left( \kappa^2 - 4 \Omega^2 \right) =0.
\label{MRI_cyl_char_eqA2}
\end{equation}
The stability criterion for this characteristic equation is given by inequality (\ref{MRI_stability_cond4}), which is the condition obtained in \cite{Balbusbook}. 

 The third case happens when $\zeta$ is such that both the $\zeta^2$ and the $\zeta$ terms are negligible. In this case, the characteristic equation yields
 \begin{equation}
\omega^2 \left( \omega^2  \frac{1}{\cos^2 q } - \kappa^2  \right) =0,
\label{MRI_cyl_char_eqA3}
\end{equation}
which is the characteristic equation of the Rayleigh shearing configuration (\ref{rs_newt_red_char_eq}), where $k_z$ is eliminated using the direction cosine and the stability criterion is (\ref{RS_criterion}). \\

In order to shed more light into the stability scenarios we will quantify the above mentioned three cases. Suppose there is a value $\zeta_\star<1$ which is the largest possible value for which both $\zeta$ and $\zeta^2$ are small enough to be neglected (i.e. the third case of the stability analysis of equation (\ref{MRI_cyl_char_eqA1}) mentioned above). Please note that we introduce this upper limit value of $\zeta$ in order to compare linear and quadratic powers of $\zeta$. Since we are interested in values of $\zeta$ that are close to zero, we can introduce this assumption without any loss of generality. For all values of $\zeta \le \zeta_\star$ the characteristic equation reduces to the Rayleigh shearing equation. The value $\zeta_\star$, in other words, is the largest value for which $\zeta,\: \zeta^2$ are effectively zero, by its definition.

For values $\zeta_\star<\zeta \le \sqrt{\zeta_\star}$ (note that the square root is larger than the number itself since $\zeta_\star<1$, and the right bound is the value such that $\zeta^2=\zeta_\star$) the linear terms in $\zeta$ do not vanish whereas the $\zeta^2$ terms can be neglected. For this interval the stability criterion is given by condition (\ref{MRI_stability_cond4}), derived in \cite{Balbus1991}.

Further increase of $\zeta$, i.e. $\zeta> \sqrt{\zeta_\star}$, implies that both the $\zeta$ and the $\zeta^2$ terms are comparable to the rest of background terms and therefore they cannot be neglected. In this case the stability condition is that given by the inequality (\ref{MRI_stability_cond2}). \\ 

Up to this point we discussed the magnitude of $\zeta$ without examining the magnitudes of the individual factors, $k^2$ and $v^2_{\rm A z}$. There is a fundamental difference between these two quantities. The former characterises the perturbation given in equation (\ref{ansatz}) and is allowed to obtain all the values within the limits that are physically meaningful, as will be discussed below. The latter is a background quantity, which corresponds to a specific axisymmetric magnetic field function for each configuration. Consequently we have the follow implication. For each $v^2_{\rm A z}$ value, which describes a physical system, we need to consider all possible $k$ values in order to make a statement regarding the stability of the system. Under the scope of the present analysis, a system is stable if all conditions are met for all possible wavenumbers. If some of the wavenumbers do not satisfy the stability condition then the system is unstable. 

\begin{figure}
	\includegraphics[width=8.5cm]{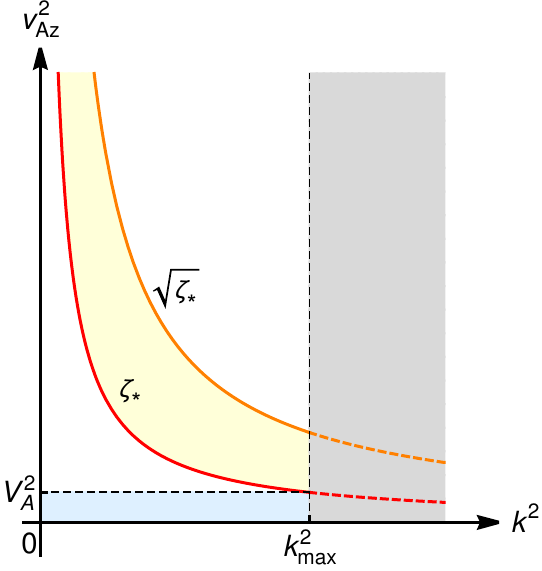}
    \caption{$v^2_{\rm A z}$ vs $k^2$ plot of $\zeta_\star$ with red and $\sqrt{\zeta_\star}$ with orange. Values of $k^2>k^2_{\rm max}$ (gray shaded region) cannot be considered since the wavenumber becomes physically impossible to exist. Consider a configuration with angular velocity $0\le\kappa^2 < 4 \Omega^2$. For any value of $v^2_{\rm A z}> V^2_{\rm A}$ it is possible to find $k$ values such that  $\zeta_\star<\zeta \le \sqrt{\zeta_\star}$  (yellow shaded region) therefore by condition (\ref{MRI_stability_cond4_b}) these configurations are unstable. However if $v^2_{\rm A z}\le V_{\rm A}^2$ (light-blue shaded region) everywhere, then necessarily $\zeta \le \zeta_\star$ and the configuration is stable, same as the case where the magnetic field was not introduced at all (see section \ref{Rayleigh_section}). }
    \label{fig1}
\end{figure}

This has the following consequence. Suppose there exists a system as the one described in section \ref{MRI} with $0\le \kappa^2<4\Omega^2$ so that the Rayleigh criterion (\ref{RS_criterion_easy}) is satisfied but the criterion (\ref{MRI_stability_cond4_b}) of MRI does not. For a given background value of $v_{Az}$, if a wavenumber value exists such that $\zeta_\star<\zeta \le \sqrt{\zeta_\star}$, then the system is unstable. As discussed in \cite{ArmitageBook, Balbusbook}, the peculiar result in this analysis stems from the fact that the zero magnetic field limit of the system is still unstable, whereas considering the same system in the context of pure hydrodynamics, as in section \ref{Rayleigh_section}, the system is stable. The resolution of this apparent physical paradox lies in the feasible range of values that the wavenumber $k$ can obtain. This is justified by the continuum hypothesis, i.e. that wavenumbers (and frequencies) of mechanical waves have some upper finite bound defined by the microscopic properties of continuous medium under consideration. Roughly, the wavelength (i.e. the inverse of wavenumber times $2 \pi$) cannot be less than the mean free path of the particles consisting the medium \citep{MorseAcoustics, TadmorBook}. Therefore, the wavenumber has an upper finite limit $k_{\rm max}$ in order to be physically possible to exist. Beyond this limit the physical system cannot be described by equations (\ref{cont_newt}), (\ref{eul_lan_newt}), and (\ref{entropy_cons_newt}) hence a different approach would be required. Given our explanation above, for certain values of $v_{Az}$ we have shown that no values of $k$ exist such that $\zeta_\star<\zeta \le \sqrt{\zeta_\star}$, therefore the system is stable. Indeed, for very small, approaching to zero, values of $v^2_{\rm A z}$ (i.e. for $v_{Az}^2 \le V_A^2$ as shown in Figure \ref{fig1}) there do not exist physically possible values of $k$, such that $\zeta_\star<\zeta \le \sqrt{\zeta_\star}$. Instead we have $\zeta \le \zeta_\star$ for the viable values of $k$. Hence, the appropriate criterion for this vanishing magnetic field limit is (\ref{RS_criterion}), as it would be if we did not introduce the magnetic field at all. Thus,  the system is stable, as expected from the hydrostatic analysis in section \ref{Rayleigh_section}. This is shown collectively in Table \ref{table1}.

On the opposite limit as the magnetic field obtains larger values  ($v_{\rm Az}^2 \rightarrow +\infty$) there are fewer and fewer wavenumbers that satisfy $\zeta_\star<\zeta \le\sqrt{\zeta_\star}$, i.e. those that fall into the yellow shaded region in Figure \ref{fig1}. This implies that in this limit the system is more stable which is in agreement with \cite{Chandra1960, Balbus1991}.

\begin{table}
	\includegraphics{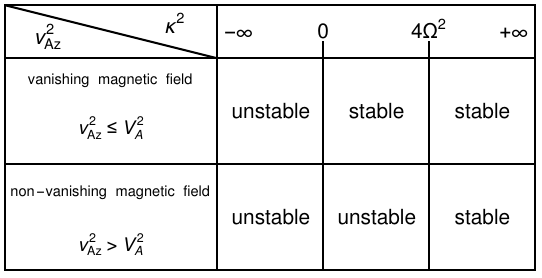}
    \caption{Stability characterisation for different cases of $\kappa^2$ and magnetic field. As mentioned previously we use $v^2_{\rm A z}$ to examine the magnetic field strength.}
    \label{table1}
\end{table}

\section{Discussion}

We used similar assumptions to \cite{Balbusbook} to derive the characteristic equation of the Rayleigh shearing instability and of the MRI in sections \ref{Rayleigh_section} and \ref{MRI}, respectively. In particular, we used the geometric optics approach to derive the characteristic equation in both of the cases mentioned above. We assumed a slow or fast variance of each of the background quantities with respect to the coordinates and time in the sense of the two timing method. Based on the assumed functional forms of our quantities, we obtained the linearised system of equations and we derived the characteristic polynomial. Additionally, our derivation did not employ the Boussinesq approximation to derive the characteristic polynomial, but the complete form of the continuity equation was used. This approach allowed us to derive the full characteristic equation which is a sixth degree polynomial in $\omega$ and it is presented in Appendix \ref{appendix}. This expression includes the terms related to the acoustic waves and to the small radial distances. By removing these terms from the full characteristic polynomial, we reached the same expression as in \cite{Balbusbook}. 

Regarding the stability characterisation mismatch of configurations that have decreasing angular velocity profiles but increasing $\Omega R^2$, we have shown that the MRI criterion is applicable if the magnetic field is above some small but finite value. Below this value such configurations are characterised by the Rayleigh shearing instability criterion, because there do not exist physically possible wavenumbers that are infinitely large. To wrap up, we have found that weak magnetic fields give rise to the MRI, however, extremely weak magnetic fields can be disregarded entirely when interested in the stability of a differentially rotating fluid as described in section \ref{Sec_rs_limit_MRI}. 

As it is obvious from section \ref{MRI}, from a strictly mathematical point of view, by taking the $k^2 v_{\rm A z}^2 \rightarrow 0$ in the MRI characteristic equation (\ref{MRI_cyl_char_eq}) we obtain the Rayleigh characteristic equation (\ref{rs_newt_red_char_eq}). However, by looking into this in more detail (see Figure \ref{fig1}), we have managed to obtain the limiting case between the MRI criterion and the Rayleigh criterion, i.e. we have found the exact regions of quantities $v_{Az}$ and $k$ where each of the stability criteria (Rayleigh or MRI) holds.

\section*{Acknowledgements}

Both authors acknowledge support from the International Hellenic University Research Scholarship.

\section{Appendix} \label{appendix}
 The full characteristic equation (along with a double $\omega=0$ root) is given by
 \begin{multline}
 \begin{aligned}
&  \left[ \omega^4 k^2 - \omega^2 k_z^2 \left( \kappa^2 + 2 k^2 v_{\rm Az}^2 \right) + k_z^4 v_{\rm A z}^2 \left( \kappa^2 - 4 \Omega^2 + k^2 v_{\rm Az}^2 \right) \right] c_{\rm s}^2 \\
& - \left\{ \omega^6 -\omega^4 \left[ \kappa^2 + k_R^2\left( v_{\rm Az}^2 + v_{\rm A\phi}^2 \right) -k_z \left(2 v_{\rm Az}^2 + v_{\rm A\phi}^2 \right]  \right) \right. \\ 
&    + \frac{\omega^2}{\rho_\bg^2}\left[  2 \Par{P_\bg}{R}\Par{P_\bg}{z} k_R k_z  +k_R^2 \left( \rho_\bg^2 k_z^2 v_{\rm Az}^2 \left(  v_{\rm Az}^2 + v_{\rm A\phi}^2 \right) -\left( \Par{P_\bg}{z} \right)^2   \right) \right. \\
& \left.  + k_z^2  \left( \rho_\bg^2 \left( v_{\rm Az}^2 \left( v_{\rm A\phi}^2 k_z^2+ \kappa^2 -4\Omega^2 \right)+v_{\rm Az}^4 k_z^2+  v_{\rm A\phi}^2 \kappa^2 \right)-\left(\Par{P_\bg}{R}\right)^2 \right)  \right] \\
&   + \frac{\omega}{\rho_\bg}  \left[ 4\Omega k_z^2 v_{\rm Az} v_{\rm A\phi} \left( k_z \Par{P_\bg}{R} - k_R \Par{P_\bg}{z} \right)  \right] \\
& \left. + \frac{1}{\rho_\bg}\left( k_z \Par{P_\bg}{R} - k_R \Par{P_\bg}{z} \right)^2 k_z^2 v_{\rm Az}^2 \right\}  \\
&- \frac{1}{R \rho_\bg^2} \left\{ {\rm i\,} \left( \omega^2 - k_z^2  v_{\rm Az}^2 \right) \left[  k_R \left(\rho_\bg^2 \left(c_{\rm s}^2 \left(\omega^2 - v_{\rm Az}^2 k_z^2\right)+\omega^2 v_{\rm Az}^2\right)  \right. \right. \right.\\
& \left. \left.  +\left(\Par{P_\bg}{z}\right)^2\right) -\Par{P_\bg}{R} \Par{P_\bg}{z} k_z+ {\rm i\,} \Par{P_\bg}{R} \rho_\bg \omega^2 \right]   \\
&    +2 \rho_\bg \omega  \Omega  v_{\rm Az} v_{\rm A\phi} k_z \left(4 \rho_\bg c_{\rm s}^2 k_z^2+{\rm i\,} \Par{P_\bg}{z} k_z-3 \rho_\bg \omega ^2\right) \\
& + \rho_\bg v_{\rm A\phi}^2 \left[ 4 \rho_\bg^2 \omega  \Omega  v_{\rm Az} v_{\rm A\phi}^3 k_z^3+2  \Par{P_\bg}{R} k_z^2 \left(v_{\rm Az}^2 k_z^2+\omega ^2\right)  \right. \\
& \left. \left. +{\rm i\,} k_R \left(v_{\rm Az}^2 k_z^2 \left(-\rho_\bg \omega ^2+2 {\rm i\,} \Par{P_\bg}{z} k_z\right)+2 {\rm i\,} \Par{P_\bg}{z} \omega ^2 k_z+\rho_\bg \omega ^4\right)\right]  \right\}  \\
& +\frac{v_{\rm A\phi}^2}{R^2 \rho_\bg^2} \left\{ \left(v_{\rm Az}^2 k_z^2+\omega^2\right) \left[\rho_\bg \omega^2  \right. \right. \\
& \left. \left. -k_z \left(\rho_\bg k_z \left(v_{\rm A\phi}^2+2 c_{\rm s}^2\right)+{\rm i\,} \Par{P_\bg}{z}\right)\right] \right\} =0,
\end{aligned}
\label{MRI_full_char_pol}
\hspace{0.92cm}
\end{multline}
where $v_{\rm A \phi}^2=\frac{B_{\bg,\phi}^2}{4 \pi \rho_\bg}$.



\end{document}